\documentclass[10pt,conference]{IEEEtran}
\IEEEoverridecommandlockouts
\usepackage{cite}
\usepackage{amsmath,amssymb,amsfonts}
\usepackage{algorithmic}
\usepackage{graphicx}
\usepackage{textcomp}
\usepackage{xcolor}
\usepackage{multirow}
\usepackage{xspace}
\usepackage[ruled,linesnumbered]{algorithm2e}

\usepackage[flushleft]{threeparttable}

\usepackage{flushend}

\def\BibTeX{{\rm B\kern-.05em{\sc i\kern-.025em b}\kern-.08em
    T\kern-.1667em\lower.7ex\hbox{E}\kern-.125emX}}

\newcommand{\framework}[0]{QuPAD\xspace}
\renewcommand{\figurename}{Figure}
\begin{document}

\title{Toward Consistent High-fidelity Quantum Learning on Unstable Devices via Efficient In-situ Calibration}

\author{\IEEEauthorblockN{
Zhirui Hu\textsuperscript{\dag,\S}, 
Robert Wolle\textsuperscript{\ddag,\S},
Mingzhen Tian\textsuperscript{\ddag,\S},
Qiang Guan\textsuperscript{\pounds},
Travis Humble\textsuperscript{\P},
Weiwen Jiang\textsuperscript{\dag,\S}}
\IEEEauthorblockA{\textsuperscript{\dag}Department of Electrical and Computer Engineering, George Mason University, VA, USA.\\
\textsuperscript{\S}Quantum Science and Engineering Center, George Mason University, VA, USA.\\
\textsuperscript{\ddag}Department of Physics \& Astronomy, George Mason University, VA, USA.\\
\textsuperscript{\pounds}Department of Computer Science, Kent State University, OH, USA.\\
\textsuperscript{\P}Oak Ridge National Laboratory, TN, USA\\
\{zhu2, wjiang8\}@gmu.edu
\vspace{-0.15in}}
}



\maketitle

\begin{abstract}

In the near-term noisy intermediate-scale quantum (NISQ) era, high noise will significantly reduce the fidelity of quantum computing.
What's worse, recent works reveal that the noise on quantum devices is not stable, that is, the noise is dynamically changing over time.
This leads to an imminent challenging problem: At run-time, is there a way to efficiently achieve a consistent high-fidelity quantum system on unstable devices?
To study this problem, we take quantum learning (a.k.a., variational quantum algorithm) as a vehicle, which has a wide range of applications, such as combinatorial optimization and machine learning.
A straightforward approach is to optimize a variational quantum circuit (VQC) with a parameter-shift approach on the target quantum device before using it; however, the optimization has an extremely high time cost,
which is not practical at run-time.
To address the pressing issue, in this paper, we proposed a novel \underline{qu}antum \underline{p}ulse-based noise \underline{ad}aptation framework, namely QuPAD.
In the proposed framework, first, we identify that the CNOT gate is the fidelity bottleneck of the conventional VQC, and we employ a more robust parameterized multi-qubit gate (i.e., Rzx gate) to replace CNOT gate.
Second, by benchmarking Rzx gate with different parameters, we build a fitting function for each coupling qubit pair, such that the deviation between the theoretic output of Rzx gate and its on-device output under a given pulse amplitude and duration can be efficiently predicted.
On top of this, an evolutionary algorithm is devised to identify the pulse amplitude and duration of each Rzx gate (i.e., calibration) and find the quantum circuits with high fidelity.
Experiments show that the runtime on quantum devices of QuPAD with 8-10 qubits is less than 15 minutes, which is up to 270$\times$ faster than the parameter-shift approach.
In addition, compared to the vanilla VQC as a baseline, QuPAD can achieve 59.33\% accuracy gain on a classification task, and average 66.34\% closer to ground state energy for molecular simulation.

\end{abstract}

\begin{IEEEkeywords}
Quantum Learning, Quantum Noise, Unstable Noise, Noise Adaptation, Pulse Calibration.
\end{IEEEkeywords}

\section{Introduction}

Quantum learning has a wide range of applications with theoretic proof of quantum speedup over classical computing\cite{yang2023device,yang2022hardware,liao2021shadow,pokharel2022demonstration}, such as molecular simulation \cite{bernal2022perspectives,kumar2023quantum,kumar2022accurate}, combinatorial optimization \cite{ranvcic2023noisy,kazi2023landscape,alam2023solving,shaydulin2023parameter}, and machine learning \cite{jiang2021co,caro2023out,tang2022dequantizing,li2023novel,wang2023qumos}.
However, such quantum speedup can dismiss when deploying to actual quantum devices. 
Specifically, as We are currently in the Noisy Intermediate-Scale Quantum (NISQ) era, the high level of noise in quantum computing is a roadblock to unleashing the power of quantum learning for real-world applications.
One promising solution is quantum error correction (QEC) \cite{shor1995scheme,gottesman1997stabilizer} for fault-tolerant quantum computing, which however requires thousands or even more physical qubits for one perfect qubit; this is not practical in the near-term NISQ systems.
Another possible solution is Quantum Error Mitigation (QEM) \cite{giurgica2020digital,endo2018practical,huggins2021virtual}; however, a recent work \cite{takagi2021fundamental} reveals the scalability issue of QEM, which has exponentially growing overhead with circuit depth.
The scalability issue is magnified in near-term quantum devices, where the noise is not stable \cite{dasgupta2022assessing}.
It, therefore, calls for a more lightweight noise suppression technique to deal with temporal varied quantum noise, called \textbf{Quantum Error Adaptation (QEA)} in this paper.

\begin{figure}[t]
\centering
\includegraphics[width=\linewidth]{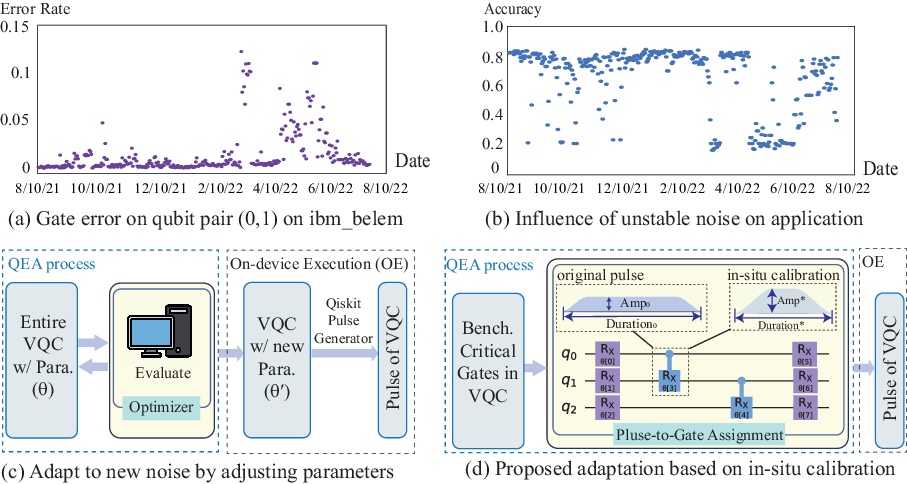}
\caption{Illustration of unstable noise, its influence, and solutions: (a) unstable noise leads to fluctuating error rates of the two-qubit gate over one year; (b) influence of unstable noise on quantum learning-based classification; (c) straightforward applying the existing technique to address unstable noise; (d) the proposed lightweight adaptation using in-situ calibration.}
\label{fig:intro}
\end{figure}

Recently, there are growing research works \cite{BetisDSN21,dasgupta2022assessing,dasgupta2022adaptive,hu2023battle,ahn2023non,sadlier2020characterization} noticed that the 
quantum noise exhibits considerable variability over time, which can result in the system fidelity being inconsistent at different times.
As an example, we trace the error rate on IBM's actual quantum processors for one entire year and tested the performance of a quantum circuit for a classification task.
Figure \ref{fig:intro}(a) show the error rate of a pair of qubits, which varies from 0.005 to 0.13 in a year.
As a result, the accuracy of the classification task changed from 0.82 to 0.2, as shown in Figure \ref{fig:intro}(b).
In realistic applications, users are usually blind to the application's performance caused by noise, since quantum vendors only provide noise data.
Therefore, it is critical to have a systematic approach to transparently ensure a stable performance of a quantum learning system.


To battle against unstable noise, a QEA process on the target quantum device is essentially needed before the on-device execution (OE), as shown in Figure \ref{fig:intro}(c)-(d).
For quantum learning applications, a straightforward QEA process is to optimize VQC with a parameter-shift approach\cite{wang2022qoc}, as shown in Figure \ref{fig:intro}(c); however, the optimization has an extremely high time cost. 
For one thing, the entire quantum circuit needs to be executed.
In addition, the optimization needs to be iteratively executed between quantum and classical computing, bringing high timing costs.
As an example, the cost of optimizing a VQC with 27 parameters on 150 samples for one epoch is about 6.75 hours.
Although there exists recent work to reduce the frequency of optimization in the QEA process by using a clustering method based on historic data \cite{hu2023battle}, it is prohibitive to conduct optimization at run-time when the level of noise changes significantly.
Therefore, a more efficient solution is needed for the QEA process.



Unlike the existing work to adapt new noise from the software level by changing the parameters in the quantum circuit, we propose to perform quantum error adaptation from the pulse control level. 
Our main hypothesis is that the unstable noise makes the pulse needing to be calibrated again to maintain high performance, instead of changing parameters only.
In this paper, we make the observation that pulses with different configurations (i.e., duration) will theoretically have the same function, but, due to control precision and interaction with the environment, the on-device results are quite different.
More importantly, for the same qubits with varied noise levels at different time points, the optimal configuration for the highest gate fidelity changes.
Therefore, instead of optimizing the parameters, we propose to perform in-situ pulse calibration and assign the pulse with the same function but with higher fidelity to the quantum gates in VQC.

Specifically, we proposed a novel quantum pulse-based noise adaptation framework, namely QuPAD, which is a two-stage optimization framework.
The first stage is performed offline: We propose a duration-aware training that takes the circuit duration into consideration when optimizing parameters in quantum circuit ansatzes; after this stage, the parameters of ansatzes will be fixed, which will be used in the second stage.
The second stage is performed online: we will establish a look-up table (LUT) to record the fitting function for each coupling qubit pair, such that the deviation between the theoretic output of target gate and its on-device output under a given pulse amplitude and duration can be efficiently predicted; based on LUT, we employ Covariance Matrix Adaptation Evolution Strategy (CMA-ES) to calibrate the pulse amplitude and duration of each Rzx gate and find the quantum circuits with high fidelity.
The calibration is performed right before executing the quantum circuit, and we call this process ``\textbf{in-situ calibration}''.

The main contributions of this paper are as follows.
\begin{itemize}
    \item We reveal the optimal pulse parameters (i.e., duration and amplitude) for the same function quantum gate changes over time because of the unstable noise in quantum devices. On top of this, we propose an in-situ pulse calibration to improve the fidelity of quantum learning.
    \item We show that the continuous cross-resonance-based gate (i.e., $R_{zx}$ gate) can provide higher fidelity than the CNOT gate, but the value of parameters in $R_{zx}$ gate will affect the duration. Correspondingly, we propose a duration-aware optimization to tune parameters with a bi-objective on minimizing the duration and maximizing accuracy.
    \item A holistic framework, namely QuPAD, is proposed to transparently optimize the quantum system implementation based on the duration-aware optimization at offline and the in-situ calibration at run-time.
\end{itemize}

Experiment results on molecular simulation and classification tasks show the duration reduction by over $3\times$ and performance improvements on  duration-aware training and \framework.  Specifically, \framework can maintain the entire performance on MNIST-6, which has 59.33\% performance improvement over villain VQC. Furthermore, the in-situ calibration method is $15\times$ to $270\times$ faster than the parameter shift method, in terms of circuit complexity.

The remainder of the paper is organized as follows. Section II provides observations and motivations. Section III and Section IV present the proposed framework and the detailed implementation. Experimental results are provided in Section V.
Related work is discussed in Section VI and conclusion remarks are given in Section VII.

\section{Preliminary, Observation and Motivation}


\subsection{Preliminary}

\noindent\textbf{Preliminary 1: Pulse-efficient circuit compilation}

\begin{figure}[t]
\centering
\includegraphics[width=0.9\linewidth]{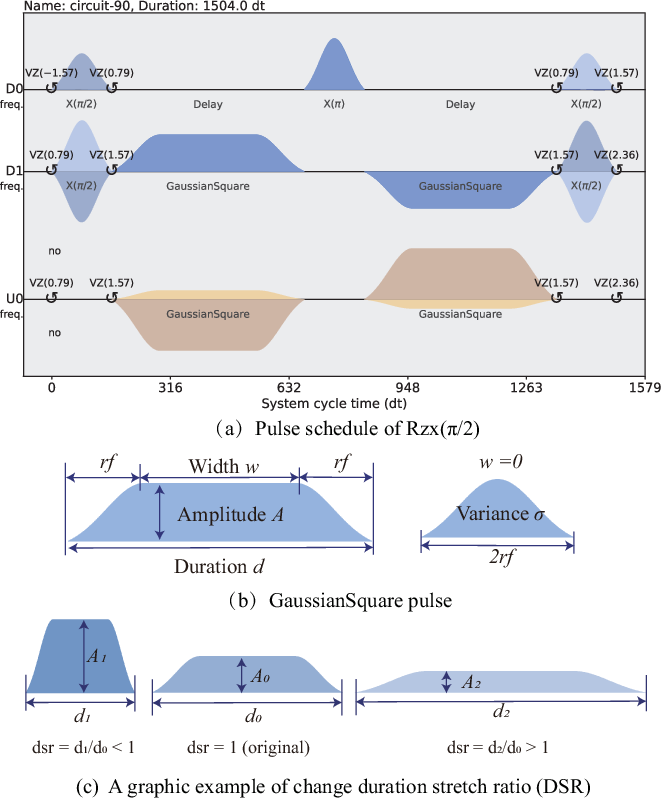}

\caption{Pulse schedule and pulse parameters, where `D0, D1' means the drive channel on qubit 0 and qubit 1, `U0' means the control channel on qubit 0.}
\label{fig:p2}
\end{figure}

\begin{figure*}[t]
\centering
\includegraphics[width=\linewidth]{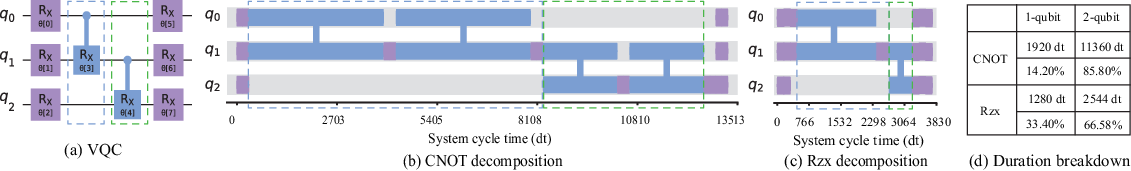}
\caption{(a) A typical unit of hardware-efficient ansatz (b)The corresponding pulse schedule with CNOT-based compiler, where purple blocks represent single-qubit gate pulse and blue blocks represent double-qubit gate pulse. (c)The corresponding pulse schedule with Rzx-based compiler. (d) The percentage of 1-qubit gate duration and 2-qubit gate duration in different circuits.}
\label{fig:timeline}
\end{figure*} 

A pulse-efficient circuit compilation workflow was proposed by IBM Quantum \cite{earnest2021pulse}. This work first decomposes a large circuit to a series of SU(4), then decomposes arbitrary SU(4) to an $Rzx$-based set, \textit{'id, u1, u2, u3, Rzx'}. 
As shown in Figure~\ref{fig:p2}(a), the approach generates the pulse of $R_{ZX}(\theta)$ by composing a series of pulses, in which the GaussianSquare pulse can be obtained by scaling the driven pulse of the CNOT gate without necessitating re-calibration. 

As illustrated in Figure~\ref{fig:p2}(b), GaussianSquare  are flat-top pulses with amplitude $A$, width $w$, risefall $rf$ and Gaussian flanks exhibiting standard deviations $\sigma$. The pulse area is then given by 
\begin{equation}
    \alpha = |A| [w+\sqrt{2 \pi} \sigma \times erf(\frac{rf}{\sqrt{2}\times \sigma} ) ].
\end{equation}

Specifically, $R_{ZX}(\theta)$ can be obtained by scaling the area of the GaussianSquare pulse according to 
\begin{equation}
\alpha(\theta) = \frac{\theta \alpha^\ast}{\pi/2}
\end{equation}
 where $\alpha^\ast$ denotes the original area of the GaussianSquare pulse of the CNOT gate.
It is worth noting that the amplitude is exclusively scaled when $w = 0$, circumventing additional calibration as the relation between $\theta$ and $w$ is linear. 
The detailed implementation of scaling GaussianSquare in CNOT pulse can be found in Appendix D of the work \cite{stenger2021simulating}.

\vspace{3pt}
\noindent\textbf{Preliminary 2: Different pulses for the same function}

The rotation angle $\theta$ of the Rzx gate is dependent on the area under the pulses, as mentioned in \cite{stenger2021simulating}. In order to maintain the function of $R_{ZX}(\theta)$, we can alter the durations and amplitudes of the GaussianSquare pulse while keeping the area fixed, which is shown in Figure~\ref{fig:p2}(c). We define the ratio of the new duration to the original duration as duration stretch ratio, (dsr), which is 
\begin{equation}
dsr = \frac{duration_{new}}{duration_{original}}
\end{equation}
and the detailed method to scale the GaussianSquare pulse is outlined in Section IV. 


\subsection{Observation and Motivation}

\noindent\textbf{Observation 1. Two-qubit gates dominate the overall duration of VQCs while compilers with different basis gates can significantly affect the duration of a two-qubit gate}

Figure~\ref{fig:timeline}(a) is a typically used ansatz, which is composed of single-qubit rotation gates and two-qubit control-rotation gates. 
In order to implement the circuit to actual quantum devices, it will be decomposed to basis gates.
The standard IBM quantum computer supports a set of basis gates \textit{\{id, u1, u2, u3, cx\}} \cite{cross2017open}, called a CNOT-based compiler.
On the other hand, the CX (or CNOT) gate can be replaced by the parameterized cross-resonance-based gate (e.g., $R_{zx}$ gate) to form a basis of \textit{\{id, u1, u2, u3, $R_{zx}$\}}, called Rzx-based compiler.

Figures \ref{fig:timeline}(b)-(c) make a comparison between these two kinds of compilations.
It is clear that the Rzx-based compilation will significantly reduce the system duration.
For the circuit in Figure~\ref{fig:timeline}(a), the cycle time is reduced from 13,513 using CNOT to 3,830 using $R_{zx}$.
More overall, as shown in Figures \ref{fig:timeline}(d), no matter which compilation to be used, the two-qubit gates will always dominate the overall duration.

\begin{table}[t]

\caption{The duration and accuracy comparison of CNOT-based compiler and $Rzx$-based compiler on IBM backend ‘ibmq\_montreal’, and the average $\bold{T_1 = 112.36 us} $ and $\bold{T_2 = 91.21 us}$. }

\renewcommand{\arraystretch}{1.3}
\tabcolsep 1 pt
\begin{tabular}{|c|c|c|cc|cc|}
\hline
\multirow{2}{*}{Benchmark$^1$} & \multirow{2}{*}{\begin{tabular}[c]{@{}c@{}}\# of\\ qubits\end{tabular}} & \multirow{2}{*}{\begin{tabular}[c]{@{}c@{}}Acc. on \\ simulator\end{tabular}} & \multicolumn{2}{c|}{CNOT-based compiler} & \multicolumn{2}{c|}{Rzx-based compiler} \\ \cline{4-7} 
 &  &  & duration(us) & Accuracy & duration(us) & Accuracy \\ \hline
MNIST-6 & 8 & 73.50\% & 57.62 & 52.50\% & 22.13 & 67.50\% \\ \hline
MNIST-8 & 10 & 70.00\% & 60.67 & 14.00\% & 22.54 & 59.00\% \\ \hline
\end{tabular}
\vspace{3pt}
\begin{tablenotes}
\scriptsize 
\item $^1$ 'MNIST-n' represents an n-class classification task.

\end{tablenotes}

\label{tab:m1}
\end{table}

\noindent\textbf{Motivation 1. Employing the continuous cross-resonance-based gate (e.g., $R_{zx}$) as the basic gate to improve fidelity.}

Based on the above observation, it motivates us to apply the $R_{zx}$ gate in compiling and designing the variational quantum circuit.
We verify the effectiveness of using $R_{zx}$ compiler by the widely used quantum learning task on MNIST sub-datasets.
Table~\ref{tab:m1} reports the results on IBM `ibmq\_montreal' quantum backend.
It clearly shows that 
$Rzx$-based compiler can reduce the duration by over $2.5\times$, which leads to a much higher accuracy on the actual quantum device, compared with the CNOT-based compiler.
However, as we can observe from the results, even using $R_{zx}$ compiler, the accuracy is still far lower than the results on perfect simulation, e.g., 6\% and 11\% accuracy drop for MNIST-6 and MNIST-8, respectively.
Therefore, we have the statement that only replace CNOT-based compiler by $R_{zx}$-based compiler is not enough, and more optimization is needed.



\noindent\textbf{Observation 2. The value of the parameter in a cross-resonance-based gate will lead to different gate duration.}

\begin{figure}[t]
\centering
\includegraphics[width=\linewidth]{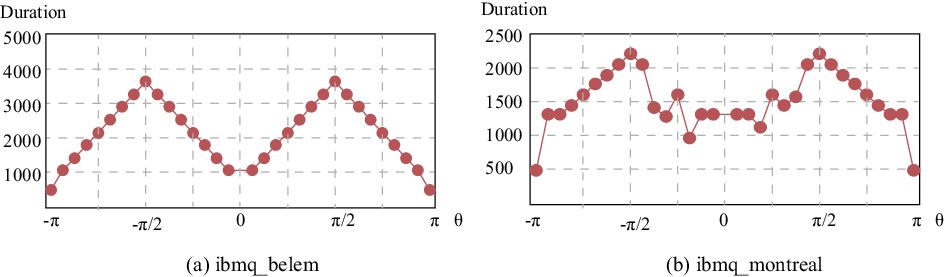}
\caption{Duration of $Rzx(\theta)$ as $\theta$ changes from $-\pi$ to $\pi$  }
\label{fig:o2}
\end{figure}

On a pair of qubits, the CNOT gate has the same duration at the pulse level; however, different gate parameters in cross-resonance-based gates may cause the corresponding pulse to have a different duration.
We test $R_{zx}(\theta)$ on two IBM Quantum backends, i.e., $ibmq\_belem$ and $ibmq\_montreal$. 
Figure~\ref{fig:o2} reports the results, where x-axis represents different value of 
$\theta$ ranging from $-\pi$ to $\pi$, and y-axis stands for the corresponding duration.
We observe common patterns of the duration variation on these two backends: Duration curves exhibit symmetry and the duration reaches its maximum when $\theta$ is around $-\frac{\pi}{2},\frac{\pi}{2}$, and minimum around $-\pi,0,\pi$.

To investigate the root cause, we found that the duration curves exhibit symmetry with respect to $\theta=0$ because 
for a pair of $\theta$ and $-\theta$, the pulse direction (upward or downward) will be inverted while the shape will not be changed.
Then, for $\theta\in (0,\pi)$, the duration would be theoretically directly proportional to the magnitude of $\theta$; however, with the objective of minimizing the duration, the compiler can decompose the gate.
For $\theta\in [\frac{\pi}{2},\pi]$, $R_{zx}(\theta)$ can be implemented by using $R_{zx}(\beta)$ where $\beta\in(0,\frac{\pi}{2})$ and single qubits gates. 
Therefore, we can observe the inflection point at $\frac{\pi}{2}$.
Due to symmetry, another inflection point is at $-\frac{\pi}{2}$.

\noindent\textbf{Motivation 2. Developing duration-aware circuit optimization to reduce the overall duration and improve fidelity.}

\begin{figure}[t]
\centering
\includegraphics[width=\linewidth]{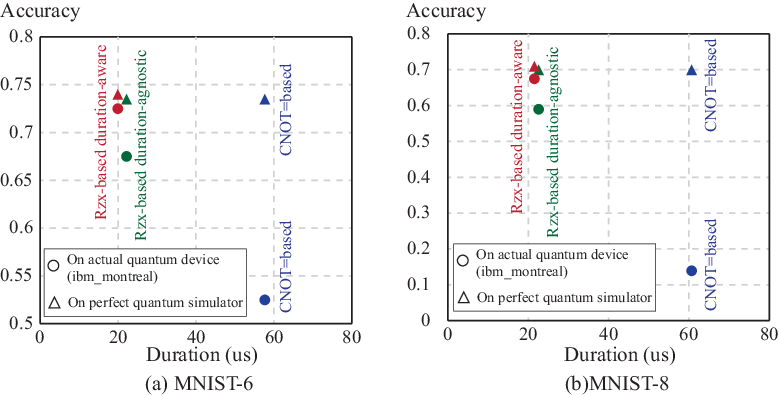}
\caption{Accuracy and fidelity comparison of different approach on perfect quantum simulator and actual quantum device (ibmq\_montreal)}
\label{fig:m3}
\end{figure} 

Based on the above observation, duration-agnostic circuit optimization may prolong circuit duration, and in turn, reduce fidelity.
Moreover, the crosstalk among qubits can cause noise \cite{ganzhorn2020benchmarking}, and it is also meant to reduce the crosstalk noise.
In the duration-aware optimization, even the overall circuit duration may not be reduced due to the schedule, optimizing one gate can reduce the time period for crosstalk.
For example, if gate $G_i$ and gate $G_j$ on different qubits are scheduled to be executed at the same time, and duration $d(G_i)<d(G_j)$.
The reduction in the duration of $G_i$ cannot reduce the overall duration, but it is still useful since it can reduce crosstalk noise.

Figure \ref{fig:m3} further show that the duration-aware optimization can indeed improve fidelity on two MNIST sub-datasets (e.g., MINST-6 with 6 classes).
In these figures, the x-axis is the overall circuit duration while the y-axis is the classification accuracy. 
The circle and triangle represent accuracy on an actual quantum computer and perfect quantum simulator.
Therefore, for each approach, the distance between circle and the triangle reflects the fidelity, i.e., the closer of these points the higher fidelity.
It is clear that using parameterized cross-resonance-based gates can reduce the duration and improve fidelity.
What's more, our proposed duration-aware approach, QuTrainer in Section III, can further reduce the duration and improve fidelity.






\noindent\textbf{Observation 3. On an actual quantum device, changing the Duration Stretch Ratios (dsrs) of a parameterized cross-resonance-based gate will affect fidelity. More importantly, the optimal dsr for maximum fidelity is varied for different noises.}

\begin{figure}[t]
\centering
\includegraphics[width=\linewidth]{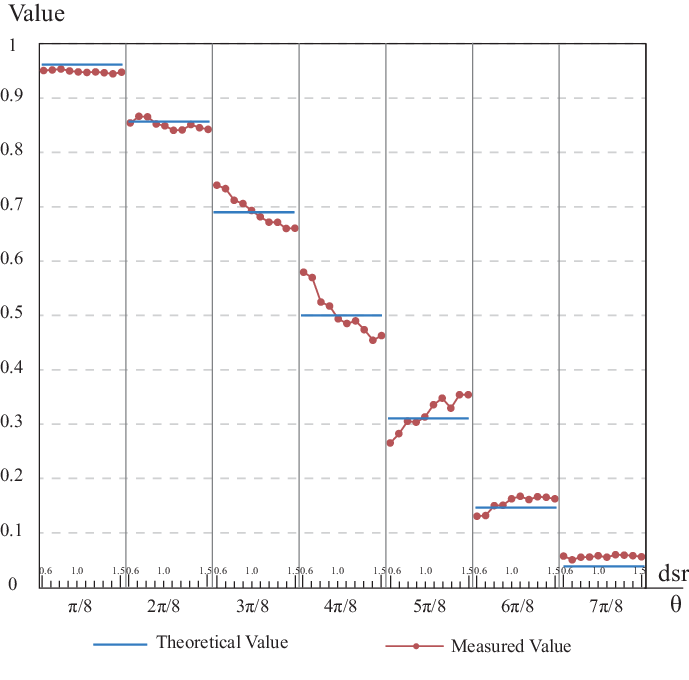}
\caption{Measured values of $|00\rangle$ with $\theta$ and duration changes on qubit pair (0,1)}
\label{fig:er2}
\end{figure}

\begin{figure}[t]
\centering
\includegraphics[width=\linewidth]{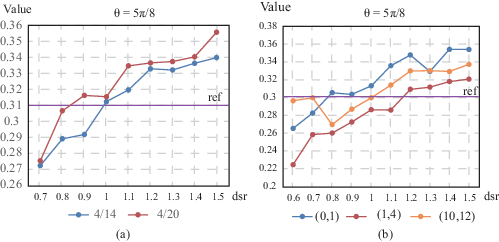}
\caption{ (a) Measured values of $|00\rangle$ on different days. (b) Measured values of $|00\rangle$ on different qubits pairs. }
\label{fig:variance}
\end{figure} 


In addition to duration, we observe another factor that affects gate fidelity.
As discussed in \textbf{Preliminary 2}, the function of $R_{zx}$ gate is determined by the area under the Gaussian-Square pulse, and we defined Duration Stretch Ratios (dsrs) to guarantee the area by changing pulse duration and amplitude.
We tested different dsrs for a set of $R_{zx}(\theta)$ on the actual quantum devices and observe that different dsrs will significantly influence fidelity when $\theta$ approaching to $\frac{\pi}{2}$.

Due to the symmetry property, we test $R_{zx}(\theta)$ with different dsrs for $\theta\in[0,\pi]$.
Figure \ref{fig:er2} reports the results, where we recorded the measured value of $|00\rangle$ when a single $R_{zx}(\theta)$ gate is placed on qubit pair $(0,1)$ as dsrs were varied from 0.6 to 1.5.
In the figure, we also plot the theoretical value as a reference, which can help us to measure the gate fidelity, that is, the closer the value is to the reference value the higher fidelity that the gate has.
We observed that when $\theta$ changes, the points of intersection are also different. This indicates that the optimal dsr (in terms of fidelity) is different by different parameter values $\theta$.
In addition, as $\theta$ approaches $\pi/2$, the variation of output value is enlarged, indicating dsr has a larger influence on fidelity.







Now, a more interesting question is whether the optimal dsr will be changed along with noise. 
Since we know that the quantum noise is not stable (see Figure \ref{fig:intro}), if the above hypothesis is true, we cannot use the same dsr.
We explore the dsr and fidelity relationship under two scenarios where noise level will change: (1) temporal variation, we test on the same pair of qubits but different dates; (2) spatial variation, we test on different pairs of qubits on the same day.
Results in these two scenarios are reported in Figure \ref{fig:variance}(a)-(b), where we can clearly see that the optimal dsr changes.

\noindent\textbf{Motivation 3. Integrating in-situ calibration in executing VQC for consistent high fidelity.}

\begin{figure}[t]
\centering
\includegraphics[width=0.9\linewidth]{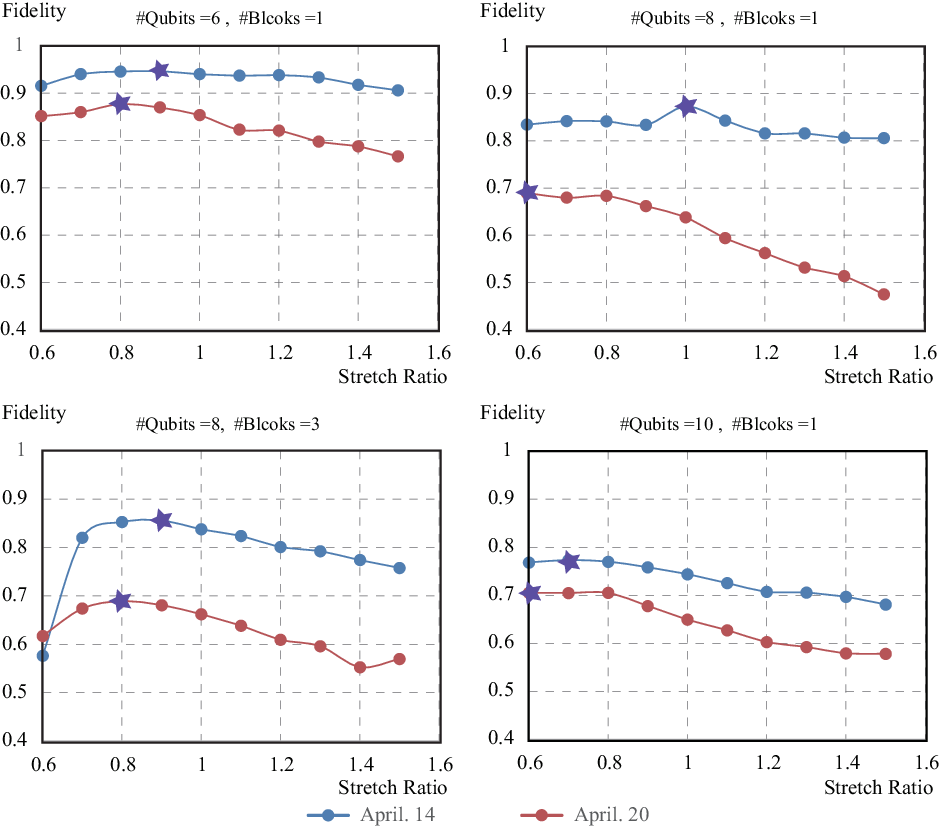}
\caption{The fidelity vs. dsr for different scales of hardware-efficient ansatz on two different days}
\label{fig:o4}
\end{figure}


The above observation motivates us that the performance of an application may be affected by dsr.
More importantly, we need to identify the best dsr at different days, since the noise changed.
We apply different scales of hardware-efficient VQC ansatz  with random parameters and we change the dsr of all $Rzx$ gates to plot the relationship between fidelity and dsr.
Figure \ref{fig:o4} report the results and verify the above statement. 
Specifically, on April 14 and 20, the optimal dsrs for the highest fidelity are different (marked as stars).

The consideration of unstable quantum noise and can affect the optimal dsrs, inspires us to carry out an in-situ calibration at run-time, which will determine the optimal dsr and in turn, determine the pulse duration and amplitude.

\begin{figure*}[!t]
\centering
\includegraphics[width=\linewidth]{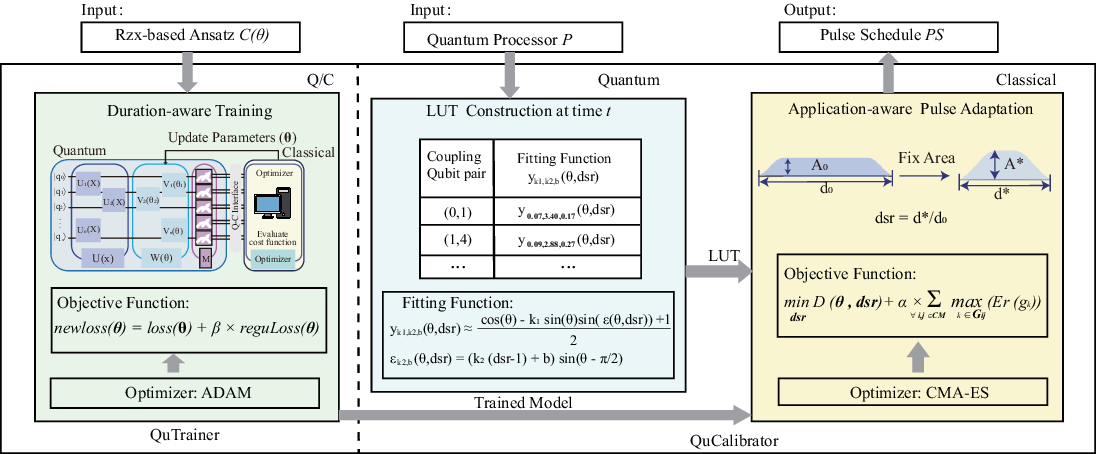}
\caption{Overview of the proposed \framework framework: (left) Duration-aware QuTrainer; (right) QuCalibrator with novel in-situ calibration}
\label{fig:framework}
\end{figure*}

\section{Methods} 

In this section, we introduce our proposed framework, designated as \framework, designed to achieve high-fidelity quantum learning. Before exploring the details, we first present a formal problem formulation:

Given an $Rzx$-based quantum learning circuit, $C(\boldsymbol{\theta} )$, where all entanglement gates are $Rzx$ gates and $\theta$ denotes the trainable parameters, and a quantum processor $Q$ at sampled time $t$, the objective is to identify the optimal parameter set $\boldsymbol{\theta}$ for circuit $C$. Additionally, we seek to adjust the pulse of each $Rzx$ gate, characterized by duration $d$ and amplitude $A$, with the goal of producing an output pulse schedule ($PS$) that maximizes the accuracy of $PS$ on the quantum processor $Q$.

\subsection{Framework overview}

Figure~\ref{fig:framework} presents an overview of our proposed two-step framework, which comprises three primary components: (1) a duration-aware training algorithm, (2) a look-up table (LUT) constructor, and (3) an application-aware pulse adaptation algorithm. The LUT constructor divides the \framework into two stages, called QuTrainer and QuCalibrator.

The goal of duration-aware training is to attain a well-trained model characterized by high accuracy and short duration. Initially, we employ $Rzx$ gates as the two-qubit  gates to construct a hardware-efficient ansatz. Subsequently, the parameters are updated based on the task performance. The output of duration-aware training is the well-trained model $C(\boldsymbol{\theta})$.

The LUT constructor is crucial for in-situ calibration. Given a quantum processor $Q$ at time $t$, we determine the fitting function for each coupling qubit pair to fit the value of $Rzx(\theta)$ to construct the LUT $T$. 

In the application-aware pulse adaptation algorithm, we use the CMA-ES-based algorithm to search the optimal duration stretch ratio $dsr$ and amplitude $A$ while maintaining a fixed area for every GaussianSquare pulse in $Rzx(\theta)$ pulse series. This adjustment aims to minimize the total circuit duration $d_c$ and the sum of the maximum gate error rates on each qubit, ultimately maximizing the fidelity of circuit $C(\boldsymbol{\theta})$ on quantum processor $P$ and time $t$.


\subsection{Duration-aware optimization}

To get a short-duration QNN with high fidelity, We can heuristically make the duration of each $Rzx(\theta)$ short. As discussed in Motivation 2, the duration of $Rzx(\theta)$ with $\theta$ near '$-\pi/2,\pi/2$' is longer than that with $\theta$ near '$-\pi,0,\pi$', which is to say, we should make the model converge to a parameter range near $-\pi,0,\pi$ during the training process. 

To achieve the tradeoff of QNN performance and short pulse duration, we rewrite the loss function as a combination of the original loss function and the regularization term as follows, which encourages some of the weights near $-\pi,0,\pi$.
\begin{equation}
\small
\begin{aligned}
reguLoss(\boldsymbol{\theta} )  =  \frac{1}{n} \sum_{i=0, g_i \in G}^{n} abs(\theta_i- round(\frac{\theta_i}{\pi})\times \pi)  \\
\label{eq:loss1_reguloss}
\end{aligned}
\end{equation} 

\begin{equation}
\small
\begin{aligned}
 newloss(\boldsymbol{\theta} )  = loss(\boldsymbol{\theta}) + \beta \times reguLoss(\boldsymbol{\theta} )
\label{eq:loss1}
\end{aligned}
\end{equation} 
where $loss(\boldsymbol{\theta})$ is the original loss function for the tasks, $G$ is the set of all $Rzx$ gates in the given circuit, $n$ is the number of $Rzx$ gates, $round()$ is a function that takes a number as input and returns the closest integer value to that number, and $\beta$ is a hyperparameter that needs to be tuned during the training process, A larger value of $\beta$  will result in a stronger regularization effect, while a smaller value will lead to a lower effect on regularization. Usually, $\beta$ is less than 0.01. Note that Equation~\ref{eq:loss1_reguloss} is a soft regularization, different from sparse training, the target of which is not to prune the weight, but to make it close to $-\pi,0,\pi$. 

 The forward propagation process can be executed on either a classical computer or an actual quantum device. If completed on a classical computer, gradient descent methods such as ADAM or SGD can be utilized for parameter updates. In contrast, if executed on a quantum computer, parameter shift methods \cite{wang2022qoc} can be employed. Throughout the parameter updating process, we heuristically converge the parameters towards fixed points with short pulse durations. Note that we can train the model using the original optimization methods with the new loss function, which will not add additional training costs.
 
\subsection{LUT construction}
We have shown in Observation 3 that the gate error rate ranges from $\theta$ and duration stretch ratio ($dsr$). (The detailed duration stretch method is shown in the next Section)
In this section, we will introduce how to build up the look-up table of $Rzx(\theta)$ error rates. 

If we initial a 2-qubit circuit as $|00\rangle$ and apply $Rzx(\theta)$ on $(q_1, q_0)$, the unitary matrix will be 

\begin{equation}
R_{Z X}(\theta) q_1, q_0 =  \exp \left(-i \frac{\theta}{2} Z \otimes X\right) 
\end{equation}

and the theoretical measured state vector should be 
\begin{equation}
 R_{zx}(\theta)|00\rangle = cos(\frac{\theta}{2})|00\rangle -i sin(\frac{\theta}{2})|01\rangle
\end{equation}
 So the theoretical measured value of $|00\rangle$ is $cos^2(\frac{\theta}{2})=\frac{1+cos(\theta)}{2}$. 
We assume the gate errors mainly originate from control errors, and the measured value of $|00\rangle$ with errors can be rewritten as 
\begin{equation}
y(\theta,dsr) =\frac{cos(\theta + \epsilon(\theta,dsr))+1}{2} 
\label{eq:error}
\end{equation}
Considering that $\epsilon(\theta,dsr)$ is close to 0 compared with 1, we can approximate $cos(\epsilon(\theta,dsr)) = 1$ and approximate the above equation as follows,
\begin{equation}
y(\theta,dsr) \approx\frac{cos(\theta) -sin(\theta)sin(\epsilon(\theta,dsr)) +1 }{2}
\label{eq:error_approx}
\end{equation}
So the fitted function can be rewritten as
\begin{equation}
y(\theta,dsr) = \frac{cos(\theta) -k_1 sin(\theta)sin(\epsilon(\theta,dsr)) +1 }{2}  \\
\label{eq:error_approx2}
\end{equation}
and the error function is 
\begin{equation}
  \epsilon(\theta,dsr) = (k_2\times(dsr-1)+b)\times sign(\theta - \frac{\pi}{2})
\end{equation}
 where $dsr$ ranges from 0.6 to 1.5, $sign()$ is a function that returns the sign of a real number, and $k_1,k_2, b$ are the parameters that need to be fitted. 

For each coupling qubit pair, we can use $n_1$ different $\theta$ and $n_2$ different $dsr$  to fit the series of curves to describe the measured value of $Rzx(\theta)$,  where $n_1$, $n_2$ are constants. The fitting accuracy increases as $n_1$ and $n_2$  increase.
 
\textbf{Complexity Analysis} Based on the aforementioned analysis, for a given circuit $C$ and a specified quantum processor $P$ at sampled time $t$, the time complexity of LUT construction is $O(\min(E_c, E_q) \times S \times n_1 \times n_2)$, where $E_c$ and $E_p$ represent the number of coupling qubit pairs in circuit $C$ and quantum processor $P$, respectively, and $S$ denotes the number of shots for the same circuit. In contrast, the time complexity of the parameter shift is $O(N_1 \times N_2 \times S \times L \times e)$, where $N_1$ corresponds to the number of tuning parameters in circuit $C$, $N_2$ signifies the number of circuit samples, $L$ represents the circuit length, and $e$ is the number of epochs required for the model to converge. It is important to note that $n_1$, $n_2$, and $e$,$N_2$ are constants unrelated to the circuit. For a circuit, we can deduce that $\min(E_c, E_q) \ll N_1  \times L$. Consequently, LUT construction proves to be more efficient than parameter shift on actual quantum processors.

\subsection{Application-aware pulse adaptation (aka. calibration)}
The final step involves adjusting the pulse of $Rzx(\theta)$ in the model according to the LUT, with the aim of enhancing the fidelity of the entire model and ensuring that the model's performance does not significantly deteriorate when executed on a actual quantum processor.
Our observations indicate that the fidelity for different circuits is related to both the circuit duration and the gate error rate.

Let $G$ represent all $Rzx$ gates in the given circuit, $\boldsymbol{CM}$ denote the set of coupling qubit pairs $(i,j)$ in the given quantum processor, and $i,j$ be the qubit indices. We propose the following objective function for this component:

\begin{equation}
\min _{\boldsymbol{dsr}} D(\boldsymbol{\theta},\boldsymbol{dsr}) + \alpha \times \sum_{\forall i,j \in \boldsymbol{CM}} \max_{k \in \boldsymbol{G_{ij}}} ( Er (g_k) )
\label{eq:loss2}
\end{equation}

Here, $D(\boldsymbol{\theta},\boldsymbol{dsr})$ signifies the total duration of the given circuit, $G_{\boldsymbol{ij}}$ is the subset of all Rzx gates on the coupling qubit pair (i,j), $Er (g_k)$ represents the gate error rate of gate $g_k$, and $\alpha$ is a hyperparameter that determines the weight of the total duration and the sum of maximum gate error on every qubit pair.

In this work, we employ the CMA-ES algorithm to update parameters. The CMA-ES algorithm is a gradient-free heuristic method, and its detailed implementation will be presented in the next section. Ultimately, we obtain the adapted pulse schedule, which will be executed on actual quantum processors.



\section{Implementation}
In this section, we will introduce how the GaussianSquare Pulse is scaled and how to leverage CMA-ES to update parameters in detail.

\subsection{Scaling the GaussianSquare Pulse}
In this paper, we divide the scaling of GaussianSquare Pulse into two steps: First, we obtain $Rzx(\theta)$ by scaling the width so that the area $\alpha(\theta) = \frac{\theta}{\pi / 2} \alpha_{cx}$, where $\alpha_{cx}$ is the area of GaussianSquare Pulse of the CNOT Gate\cite{stenger2021simulating}. We implement Qiskit's functions \textit{EchoRZXWeylDecomposition()} and \textit{RZXCalibrationBuilderNoEcho()} to achieve them.

\begin{table*}[t]

\renewcommand{\arraystretch}{1.3}
\tabcolsep 4 pt
\centering
\caption{The Comparison of \framework with a baseline on LiH and MNIST-6.}
\begin{tabular}{|c|c|ccc|ccccc|ccccc|}
\hline
Tasks & Reference & \multicolumn{3}{c|}{VQE/VQC} & \multicolumn{5}{c|}{QuTrainer} & \multicolumn{5}{c|}{QuPad} \\ \hline
\multicolumn{1}{|l|}{} & \begin{tabular}[c]{@{}c@{}}Ground\\ Energy\end{tabular} & \begin{tabular}[c]{@{}c@{}}Duration\\ (us)\end{tabular} & Energy & \begin{tabular}[c]{@{}c@{}}Energy\\  Gap\end{tabular} & \begin{tabular}[c]{@{}c@{}}Duration\\ (us)\end{tabular} & \begin{tabular}[c]{@{}c@{}}Dur. \\ Red.\end{tabular} & Energy & \begin{tabular}[c]{@{}c@{}}Energy\\  Gap\end{tabular} & \begin{tabular}[c]{@{}c@{}}Improve\\ vs.VQE\end{tabular} & \begin{tabular}[c]{@{}c@{}}Duration\\ (us)\end{tabular} & \begin{tabular}[c]{@{}c@{}}Dur.\\  Red.\end{tabular} & Energy & \begin{tabular}[c]{@{}c@{}}Energy\\  Gap\end{tabular} & \begin{tabular}[c]{@{}c@{}}Improve\\ vs.VQE\end{tabular} \\ \hline
LiH-0.4 & -6.52 & 7.53 & -6.32 & -0.2 & 2.57 & 2.93$\times$ & -6.4 & -0.12 & 40.00\% & 2.49 & 3.02$\times$ & -6.42 & -0.1 & \textbf{50.00\%} \\ \hline
LiH-1.2 & -7.84 & 7.53 & -7.52 & -0.32 & 3.03 & 2.49$\times$ & -7.67 & -0.17 & 46.88\% & 2.88 & 2.62$\times$ & -7.7 & -0.14 & \textbf{56.25\%} \\ \hline
LiH-2.0 & -7.83 & 7.53 & -7.5  & -0.33 & 2.76 & 2.73$\times$ & -7.71 & -0.12 & 63.64\% & 2.83 & 2.66$\times$ & -7.74 & -0.09 & \textbf{72.73\%} \\ \hline
Avg. & - & - & - & -0.29 & -   & - & - & -0.12 & 57.56\% & - & - & - & -0.10 & \textbf{66.34\%} \\ \hline
\multicolumn{1}{|l|}{} & \begin{tabular}[c]{@{}c@{}}Acc.\\ Simulation\end{tabular} & \begin{tabular}[c]{@{}c@{}}Duration\\ (us)\end{tabular} & Acc. & \begin{tabular}[c]{@{}c@{}}Acc.\\ Drop\end{tabular} & \begin{tabular}[c]{@{}c@{}}Duration\\ (us)\end{tabular} & \begin{tabular}[c]{@{}c@{}}Dur. \\ Red.\end{tabular} & Acc. & \begin{tabular}[c]{@{}c@{}}Acc.\\ Drop\end{tabular} & \begin{tabular}[c]{@{}c@{}}Improve\\ vs.VQC\end{tabular} & \begin{tabular}[c]{@{}c@{}}Duration\\ (us)\end{tabular} & \begin{tabular}[c]{@{}c@{}}Dur. \\ Red.\end{tabular} & Acc. & \begin{tabular}[c]{@{}c@{}}Acc.\\ Drop\end{tabular} & \begin{tabular}[c]{@{}c@{}}Improve\\ vs.VQC\end{tabular} \\ \hline
MNIST-6 & 76\% & 52.1 & 17.33\% & 58.67\% & 17.38 & 3$\times$ & 58.67\% & 17.33\% & 41.34\% & 15.63 & 3.33$\times$ & 76.66\% & 0.66\% & \textbf{59.33\%} \\ \hline
\end{tabular}
\label{tab:main results}
\end{table*}

\begin{algorithm}[t]
\caption{CMA-ES based dsr search}
\label{algorithm:cma-es}
\KwData{look-up table $LUT$, circuit $C$, initial stretch ratio for all coupling qubit pairs $\mathbf{dsr_0}$, loss function $f$ (Equation~\ref{eq:loss2}),
 the maximum number of generations $G$, the size of population $K$}
\KwResult{Optimal duration stretch ratio $\mathbf{dsr'}$}

$\mathbf{dsr} \leftarrow \mathbf{dsr_0}$ // Initialization optimizing variable

Initialize($mu, \sigma, M$) // Initialize state variables, mean $mu$, variance $\sigma$ and covariance matrix $M$

\For{$g \leftarrow 1$ \KwTo $G$}{
\For{$k \leftarrow 1$ \KwTo $K$}{
$\mathbf{dsr_k}$ = sampling\_dsr\ ($\mathbf{dsr},mu, \sigma, M$)

 $loss_k^g \leftarrow$ Evaluate $f(LUT,C,\mathbf{dsr_k})$
}
Update ($\mathbf{dsr}$) // $dsr$ with minimum loss.

Update($mu, \sigma, M$) // Update state variables
}

\end{algorithm}

The second step involves adapting the amplitude and duration so that the area of GaussianSquare Pulse can be fixed, in order to achieve $Rzx(\theta)$. Recall that the GaussianSquare Pulse function in Qiskit is defined as follows:
 
 \begin{equation}
 \small
f^{\prime}(x)= \begin{cases}\exp \left(-\frac{1}{2} \frac{(x-\text { rf })^2}{\sigma^2}\right) & x< rf  \\ 1 &  rf \leq x< rf +  w  \\ \exp \left(-\frac{1}{2} \frac{(x-( rf+w ))^2}{ \sigma ^2}\right) &  rf+ w \leq x\end{cases}
\end{equation}

 \begin{equation}
 \small
\begin{aligned}
& f(x)=\mathrm{A} \times \frac{f^{\prime}(x)-f^{\prime}(-1)}{1-f^{\prime}(-1)} = A \times g(x), \quad 0 \leq x< durat
\end{aligned}
\end{equation}
where $w$ is width, $rf$ is risefall and $\sigma$ defines the shape of gaussian function. We define $f_0(x)$ as the original function and $f_1(x)$ as the scaled function. To keep the area fixed, let 
 \begin{equation}
\int_{0}^{durat_0} f_0(x)dx = \int_{0}^{durat_1} f_1(x)dx 
\end{equation}
We define $dsr$ as the stretch ratio of duration. Considering the AWGs can only generate pulses with a duration that is a multiple of $m = 16\ dt$, we will get 
\begin{equation}
durat_1 = [\frac{durat_0 \times dsr}{m}]\times m \ dt
\end{equation} 
and the round dsr becomes 
$rdsr = \frac{durat_1}{durat_0}$
So the new width, rise-fall, $\sigma$ and amplitude of $g_1(x)$ will become
\begin{equation}
\begin{aligned}
& w_1 = rdsr \times w_0,\\
& rf_1 = rdsr \times rf_0,\\
& \sigma_1 = rdsr \times \sigma_0,\\
&A_1 = A_0 \times \frac{\int_{0}^{durat_0} g_0(x)dx}{\int_{0}^{durat_1} g_1(x)dx}
\end{aligned}
\end{equation}



\subsection{CMA-ES based dsr search}

Covariance Matrix Adaptation Evolution Strategy (CMA-ES)\cite{hansen2001completely} is a derivative-free optimization algorithm, which has also been used for searching hyperparameters of deep neural networks\cite{loshchilov2016cma}. The objective function is shown in Equation \ref{eq:loss2}, where $D(\boldsymbol{\theta},\boldsymbol{dsr})$ is obtained from the pulse generator in Qiskit. In this component, $\boldsymbol{\theta}$ is fixed, and our goal is to find the approximate optimal $\mathbf{dsr}$ to minimize the objective function.

To improve efficiency, we reduce the search space by setting the $dsr$ of $Rzx$ gates on the same coupling qubit pair to the same value. Let $dsr_{ij}$ represent the $dsr$ of the Rzx gate on the coupling qubit pair $(i,j)$. Consequently, the search space becomes $\mathbf{dsr} = [dsr_{ij}\ \forall i,j \in \mathbf{S}]$, where $\mathbf{S}$ is an application-related subset of the set of all possible coupling qubit pairs $\mathbf{CM}$.

The pseudocode for this approach is shown in Algorithm~\ref{algorithm:cma-es}. Ultimately, we obtain the optimized $\mathbf{dsr'}$.


\section{Experimental Evaluation}

\subsection{Experiments setup}


\textbf{Benchmark.}
In this work, we evaluate our proposed framework on two different tasks: (1) Machine learning classification task using VQC ansatzes on the sub-dataset of MNIST, for example, MNIST-6 indicating the classification with 6 classes from MNIST which uses 8 qubits.
We use 150 samples from the dataset to perform the test on IBM quantum compupters.
(2) Simulating Molecules using VQE ansatzes, where we evaluate the ground state energy for LiH at various interatomic distances. The ansatz contains two layers, and each layer has $8$  $Rx(\theta)$ and $8$ $Rzx(\theta)$ gates.

\textbf{Environment Setting.}
In QuTrainer, we use the `Adam' optimizer with 8 epochs for the ML task and `COBYLA' with 300 iterations for Molecules Simulation. 
In QuCalibrator, we employ two 27-qubit IBM Quantum backends, including `ibmq\_kolkata' and `ibmq\_montreal' (note: the results on `ibmq\_montreal' are obtained before its retirement).
All simulation experiments are conducted on an Intel Core i5-11300H (3.10GHz, 16GB RAM) computer. The machine learning classification task is implemented on Torchquantum\cite{hanruiwang2022quantumnas},
while other tasks are based on IBM Qiskit software.

\textbf{Baseline.} We compare the proposed QuPAD framework with Vallina VQC, which is trained using the duration-agnostic optimizer, which is set as the baseline.
For QuPAD, we obtain the results from the entire framework, and we also involve the results of using QuTrainer only.
All models are first trained in a noise-free environment.


\subsection{Main Results}

\begin{figure}[t]
\centering
\includegraphics[width=\linewidth]{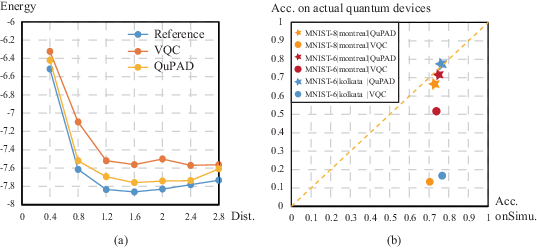}
\caption{ Evaluation of fidelity. (a) The energy gap for VQE (b) The accuracy gap between simulation and actual quantum devices}
\label{fig:vqevis}
\end{figure} 



Table~\ref{tab:main results} reports the experimental results on  simulating LiH using VQE and performing classifications on MNIST-6 using VQC.
Under the column ``Reference'', the `ground energy' for VQE is the solution obtained by SciPy minimizer; while `Simulation Acc.' for VQC is the results obtained by noise-free simulation.
The ``Energy Gap'' and ``Acc. Drop'' of QuTrainer and \framework are the difference between the reference value and the obtained value from actual noisy quantum devices. 
Column `vs. Baseline' shows the improvement over baseline. 

There are two observations. First, the performance of QuTrainer is always better than VQC. The improvement is mainly from the replacement of CNOT and the duration-aware training, which significantly reduce the duration and improve the fidelity of overall circuits.
Specifically, we can observe 2.49$\times$ to $3.00\times$ reduction in circuit length, as a result, QuTrainer can achieve solutions that are 57.56\% on average closer to the ground state energy on average and 41.34\% improvement on classification accuracy.
Then, with the in-situ calibration, the proposed \framework can further improve the system fidelity.
Specifically, for MNIST-6, it further achieves 17.99\% accuracy gain. 
Based on the above observations, we can verify the effectiveness of \framework on different tasks.

\subsection{Development Fidelity on Actual Quantum Devices}

Figure \ref{fig:vqevis} further plots the reference results and results obtained by QuPAD to investigate the fidelity.
In Figure \ref{fig:vqevis}, the x-axis is for different distances and the y-axis is for energy.
We found that the curve of \framework is closer to the reference curve in Figure~\ref{fig:vqevis}(a).
In addition, we compare the VQC on MNIST dataset on different quantum backends in Figure \ref{fig:vqevis}(b).
The x-axis is the simulation accuracy and the y-axis is the accuracy of the actual quantum device (i.e., on-device accuracy).
Ideally, if the point is on the dashed line, then we have the on-device equal to simulation accuracy, indicating a high fidelity for the quantum learning task.
From this figure, we can see that for villian VQC, although the simulation accuracy can approach 0.8, its on-device accuracy is less than 0.2.
On the other hand, QuPAD can provide solutions with much higher fidelity.
These results show that \framework can maintain fidelity while the circuits are deployed to noisy actual quantum devices for different tasks.

\subsection{Breakdown Evaluation}

\begin{table}[t]
\renewcommand{\arraystretch}{1.3}
\tabcolsep 1 pt
\caption{ accuracy comparison on ibmq\_kolkata}

\begin{tabular}{|c|ccccc|}
\hline
Task & method & duration(us) & \begin{tabular}[c]{@{}c@{}}reduction\\ (vs. VQC)\end{tabular} & \begin{tabular}[c]{@{}c@{}}Acc./Energy\\ (ibmq\_kolkata)\end{tabular} & \begin{tabular}[c]{@{}c@{}}improve\\ (vs. VQC)\end{tabular} \\ \hline
\multirow{4}{*}{\begin{tabular}[c]{@{}c@{}c@{}}VQE\\ LiH\\ dist =0.8\\Ref.$^2$=-7.61\end{tabular}}
& VQC & 7.53 & 1.00$\times$ & -7.10 & - \\ \cline{2-6} 
 & Rzx Comp. & 3.15 & 2.39 $\times$ & -7.40 & 60.00\% \\ \cline{2-6} 
 & QuTrainer & 2.79 & 2.70 $\times$& -7.42 & 62.65\% \\ \cline{2-6} 
 & QuPAD & 2.48 & 3.04 $\times$ & \textbf{-7.52} & \textbf{81.71\%} \\ \hline

\multirow{4}{*}{\begin{tabular}[c]{@{}c@{}c@{}}Classification\\ MNIST-6\\ Ref.$^1$=76\% \end{tabular}}
& VQC & 52.10 & 1.00$\times$ & 17.33\% & - \\ \cline{2-6} 
& Rzx Comp. & 19.48 & 2.67$\times$ & 53.33\% & 36.00\% \\ \cline{2-6} 
& QuTrainer & 17.38 & 3.00$\times$ & 58.67\% & 41.34\% \\ \cline{2-6} 
& QuPAD & 15.63 & 3.33$\times$ & \textbf{76.66\%} & \textbf{59.33\%} \\ \hline
 
\end{tabular}

\begin{tablenotes}
\scriptsize 
\item $^1$ The reference accuracy is obtained on the noise-free simulator with the objective of maximizing accuracy.
\item $^2$ The reference energy is obtained on the noise-free simulator by minimum eigen solver.

\end{tablenotes}

\label{tab:Breakdown}
\end{table}

Now, we want to explore how the overall improvement obtained by QuPAD, and we record the results during the optimization steps
for different methods VQE with $distance = 0.8$ and VQC on MNIST-6.
Table~\ref{tab:Breakdown} reports the results. 
The first step is to change the CNOT-based compiler to an Rzx-based compiler only and perform duration agnostic optimization.
From Table~\ref{tab:Breakdown}, we observe that the $Rzx$-based compilation can reduce the pulse duration by more than 2.5$\times$ compared to the CNOT-based compilation, while also enhancing the performance. 
The offline duration-aware QuTrainer can further obtain 2.65\% closer to ground state energy and 5.34\% accuracy gain on the actual quantum devices.
The results verify that the reduction of crosstalk duration can improve fidelity.
Finally, by integrating the in-situ calibration, QuPAD can further improve the results by 19.09\% on energy and 17.99\% on accuracy.
These results reflect that jointly applying the duration-aware optimization and in-situ calibration can obtain high fidelity for quantum applications on noisy quantum devices.

\subsection{Efficiency Evaluation}

\begin{table}[t]
\caption{ Estimated Time cost  for different benchmarks executing on IBM quantum processor.}
\centering
\renewcommand{\arraystretch}{1.3}
\tabcolsep 3 pt
\begin{tabular}{|ccccc|}
\hline
benchmark          & VQE-LiH & MNIST-4 & MNIST-6 & MNIST-8   \\ \hline
\# of Coupling pairs & 7  & 3       & 7      & 9            \\ \hline
\# of 2-qubit Gates  & 14 & 27       & 216      & 243            \\ \hline

Time cost(\framework)   & 13.5m        & 4.5m      & 13.5m     & 13.5m       \\ \hline
Time cost(Param shift\cite{wang2022qoc})      & 3.5h      & 6.75h      & 54h     & 60.75h      \\ \hline
Speedup   & 15.56× & 90×      & 240×     & 270×     \\ \hline
\end{tabular}


\label{tab:qnntime}

\end{table}

We compare the time cost of executing different scale benchmarks on quantum processors during both the LUT construction process and the parameter shift process, as shown in Table~\ref{tab:qnntime}.  
Note that for QuPAD we record the elapsed time for in-situ calibration time for comparison, and for parameter shift, we test the elapsed time to shift one qubits gate, and we calculate the overall time by-product to the total number of gates.
we observe that the time cost of LUT construction is significantly shorter than that of parameter shift on actual quantum devices. 
For all datasets, QuPAD can complete the quantum noise adaptation process within 14 minutes, while parameter shift can use over 50 hours for MNIST-6 and MNIST-8 with over 200 two-qubit gates. 
Even for the small-scale circuits, that is VQE and MNIST-4 with 14 and 27 two-qubit gates, parameter shift needs 3.5 hours and 6.75 hours, respectively.
Moreover, the speedup becomes more remarkable as the number of shifting parameters increases. Additionally, if we update different benchmarks simultaneously, we only need to establish the LUT once, which further reduces the runtime on the actual quantum processor.

\subsection{Ablation Study: QuCalibrator}

\begin{figure}[t]
\centering
\includegraphics[width=\linewidth]{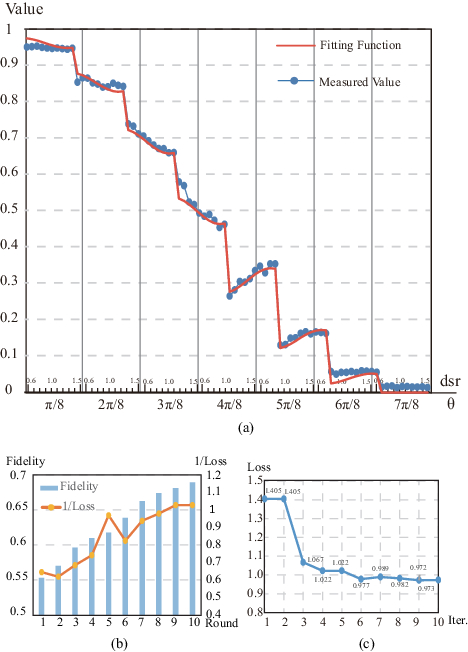}
\caption{Evaluation of QuCalibrator: (a) Fitting function on qubit pair (0,1) vs. the measured value. (b) Loss (Equation~\ref{eq:loss2}) vs. Fidelity on VQE ansatz (sorted by fidelity). (c) Loss (Equation~\ref{eq:loss2}) for each iteration of CMA-ES on VQE ansatz.}
\label{fig:adaptor}
\end{figure}

\textbf{Evaluation of  LUT construction.}
To verify whether Equation~\ref{eq:error_approx} can accurately describe the gate error as $\theta$ and $dsr$ vary, we compared the measured value of $Rzx(\theta)$ on the state of $|00\rangle$ on the qubit pair (0,1) and the fitting function in Equation~\ref{eq:error_approx}. The comparison is shown in Figure~\ref{fig:adaptor}(a). From this figure, we observe that the difference between the fitting function and the original data is small in most intervals. This demonstrates the high-accurate approximability of the proposed fitting function to the actual results under noisy quantum devices.


\textbf{Evaluation of application-aware loss function.}   
To demonstrate that the combination of normalized circuit duration and the sum of gate error in Equation~\ref{fig:adaptor}(b) is an indicator of the fidelity of the given circuit,
we randomly initialize the parameters $\boldsymbol{\theta}$ and $\boldsymbol{dsr}$ in the ansatz of VQE and simultaneously record the values of fidelity and the reciprocal of the loss. These values are sorted according to fidelity in ascending order and shown in Figure~\ref{fig:adaptor}(b). Particularly, we set $\alpha = 10$.
From Figure~\ref{fig:adaptor}(b), we observe that the change in fidelity and the change in $\frac{1}{loss}$ have a similar trend, showing that Equation~\ref{eq:loss2} can accurately indicate the variance of fidelity. 

\textbf{Evaluation of CMA-ES.} 
In order to illustrate the effectiveness of CMA-ES, we recorded the loss decrease at each iteration, as shown in Figure~\ref{fig:adaptor}(c). From this, we can see that the loss decreases sharply at the third iteration and gradually becomes stable at the 6th iteration. This indicates that after the second iteration of random sampling and variance calculation, the $\boldsymbol{dsr}$ can indeed be updated in the direction of decreasing loss, and the step size is appropriate. This demonstrates the efficiency of the CMA-ES algorithm in optimizing the loss function and finding the appropriate $\boldsymbol{dsr}$.




\section{Related Work}
\textbf{Fluctuating Noise and Methods} Several studies have addressed the issue of fluctuating noise. Dasgupta et al. \cite{dasgupta2021stability} evaluated the stability of NISQ devices using multiple metrics to characterize stability. In subsequent works \cite{dasgupta2022characterizing, dasgupta2022assessing}, they defined the Hellinger distance to measure the discrepancy between ideal states and noisy measurements. These studies emphasize the importance of replicating results during long-term execution on  quantum device. Hu et al. \cite{hu2023battle} highlighted that QNN accuracy fluctuates with noise, necessitating re-adaptation. To mitigate the impact of such noise, they proposed a cluster-based algorithm to identify different noise patterns and reuse previously obtained models with similar patterns. However, the storage requirement for these models will continuously grow. Yamamoto et al. \cite{yamamoto2022error} introduced a purification-based quantum error mitigation method, capable of mitigating systematic errors and recovering results from unknown fluctuating noise. Nonetheless, quantum devices experience various types of noise, so it is hard to mitigate all fluctuation noise by such mitigation.

\textbf{Pulse-level optimization for VQA} Liang et al. \cite{liang2022pan, liang2022variational} were among the first to employ pulse parameter shift techniques to address VQA problems. However, using gradient-based algorithms to adjust pulse parameters on actual machines is challenging, and the circuit size is difficult to scale, limiting the complexity of solvable problems. Recently, Chadwick et al. \cite{chadwick2023efficient} proposed a coordinate-based method for rapidly adjusting pulse parameters to achieve high fidelity. However, this approach targets single gates rather than application-level quantum circuits.


\section{Conclusion}

We presented QuPAD, a holistic framework that optimizes quantum systems by incorporating duration-aware optimization offline and in-situ calibration at run-time to improve the fidelity of quantum learning. 
We reveal optimal pulse parameters (i.e., duration and amplitude) for the same function quantum gate to change over time.
Motivated by this, an efficient in-situ calibration is devised to address the issue of unstable noise in quantum devices.
Experimental results demonstrate significant improvements in both duration reduction and performance
, demonstrating the effectiveness and efficiency of the proposed framework. 
Specifically, QuPAD achieves over 3$\times$ duration reduction and performance improvements over villain VQC. 
As a result, it can obtain an accuracy gain of 59.33\% over the original VQC on the MNIST dataset.
In addition, the in-situ calibration process provides a speedup of 15$\times$- 70$\times$ compared to the parameter shift method.

\section*{Acknowledgment}
This work is supported in part by Mason’s Office of Research Innovation and Economic Impact (ORIEI), Center for Trusted, Accelerated, and Secure Computing \& Communication (C-TASC), and Quantum Science and Engineering Center (QSEC). This work is also supported in part by the NSF, 2238734, 2230111, 2217021, 2212465, and 1908159. 
The research used IBM Quantum resources from the Oak Ridge Leadership Computing Facility at the Oak Ridge National Laboratory, which is supported by the Department of Energy, Office of Science, Early Career Research Program under Contract No. DE-AC05-00OR22725.  We thank Thomas Hanrui Wang \& Zhiding Liang for the discussion about VQE.

\clearpage

\bibliographystyle{ieeetr}
\bibliography{bibliography,ref}
\end{document}